\documentclass[12pt]{article}
\usepackage{epsfig}

\newcommand{\Z}{{Z \!\!\! Z}}
\newcommand{\beqn}{\begin{eqnarray}}
\newcommand{\eeqn}{\end{eqnarray}}
\newcommand{\eq}[1]{(\ref{#1})}

\newcommand{\dual}{\mbox{}^{\ast}}

\newcommand{\mon}{{\mathrm{mon}}}
\newcommand{\mix}{{\mathrm{mix}}}
\newcommand{\reg}{{\mathrm{reg}}}

\newcommand{\itep}
{~\vspace{-1.5cm}

\begin{flushright}
{\large LU-ITP 2002/023}\\
{\large KANAZAWA-02-35}\\
{\large ITEP-LAT/2002-22}\\
{\large HU-EP-02-54}
\end{flushright}
\vspace{1.0cm}}

\begin{document}

\thispagestyle{empty}

\baselineskip=14pt

\begin{center}

\itep

{\large\bf
More on String Breaking
in the $\mathbf{3D}$ Abelian Higgs Model: \\
\vspace{2mm}
the Photon Propagator}

\vskip 1.0cm {\large
M.~N.~Chernodub$^{a,b}$, E.--M.~Ilgenfritz$^c$ and A.~Schiller$^d$}\\

\vspace{.4cm}

{ \it
$^a$ ITEP, B. Cheremushkinskaya 25, Moscow, 117259, Russia

\vspace{0.3cm}

$^b$ Institute for Theoretical Physics, Kanazawa University,\\
Kanazawa 920-1192, Japan

\vspace{0.3cm}

$^c$ Institut f\"ur Physik, Humboldt--Universit\"at zu Berlin,
D-10115 Berlin, Germany

\vspace{0.3cm}

$^d$ Institut f\"ur Theoretische Physik and NTZ, Universit\"at
Leipzig,\\ D-04109 Leipzig, Germany
}

\end{center}

\begin{abstract}
We study the Landau gauge photon propagator in the
three--dimensional Abelian Higgs model with compact gauge field and
fundamentally charged matter in the London limit.
The total gauge field is split into singular and regular parts.
On the confinement side of the string breaking crossover
the momentum dependence of the total propagator is characterized
by an anomalous dimension similarly to $3D$ compact QED.
At the crossover and throughout the Higgs region the anomalous dimension
disappears.
This result perfectly agrees with recent observations that the
monopole--antimonopole plasma leads to nonzero anomalous dimension and
the presence
of the matter fields causes monopole pairing into dipole bound states.
The Yukawa mass characterizing the propagator part from regular gauge
fields is non-vanishing at the Higgs side
and coincides with the mass found for the total propagator.
The regular gauge field without anomalous dimension becomes massless
at the crossover and in the confinement region.
\end{abstract}

\newpage

\section{Introduction}

Nowadays, the interest in the lattice Abelian Higgs model with
compact gauge field (cAHM) in three dimensions has grown because
of its relation to high energy
physics~\cite{Fradkin,EinhornSavit} and its applications in
condensed matter physics~\cite{AnomalousMatter}.

The compactness of the gauge field leads to the presence of monopoles
which are instanton--like excitations in three space--time dimensions.
Being in the plasma state the monopoles and antimonopoles
of this theory guarantee linear
confinement of electrically charged test particles~\cite{Polyakov}.
They are forming an oppositely charged double sheet along the minimal
surface spanned by the Wilson loop (i.e. the trajectories of the heavy charges).
Due to screening, the free energy of the surface increases proportionally
to the area of the surface such that an area law for the Wilson loop
emerges.

Within the confinement region of cAHM corresponding
to small couplings of matter fields to gauge fields
the monopole--antimonopole plasma state is still realized. As the hopping
parameter increases, the system enters the Higgs region where
monopoles and antimonopoles become bound into magnetically neutral dipoles. This
scenario has been demonstrated in a preceding paper~\cite{CIS5} and
it has been related to the phenomenon of string breaking.
Indeed, when monopole pair formation occurs, this results in the
breakdown of linear confinement at large distances. Usually this is
interpreted in an alternative way saying that dynamical
matter fields in the same representation as the external test
charges break the confining string by screening the charge of the
latter. This argument is applied, irrespective whether the
dynamical matter field is fermionic (the quarks in QCD) or bosonic
(the Higgs particles in our case). Let us stress here the other
point of view, according to which the monopole mechanism of
confinement is changed in a way to produce a different form of the
inter--particle potential. At large separations $R$ of the charges
the string tension should be absent. However, if $R \ll
R_{\mathrm{br}}$ (where $R_{\mathrm{br}}$ is the characteristic
string breaking distance) the test charges are able to recognize
individual monopoles even if they are bound in dipoles.
Therefore, the monopole and antimonopole
fields may induce a piecewise linearly rising potential.
These simple considerations can be made more rigorous by
analytical calculations~\cite{DipoleGas} for a gas of dipoles with
small magnetic moments.

In order to destroy the linearly rising potential within cAHM$_3$,
the coupling between Higgs and gauge fields
must be sufficiently strong. It would be tempting to associate the onset
of string breaking with a phase transition between confinement and
Higgs phases. However, it was numerically shown~\cite{CIS5} that
in the London limit of cAHM$_3$ the string breaking happens in a
region of the phase diagram where a first or second order phase
transition can definitely be excluded. In the present paper we will
call this the ``string breaking crossover'', but one should keep in
mind that a thorough reconstruction of the monopole configurations is
accompanying this. However, the possibility~\cite{AnomalousMatter}
that string breaking (and monopole pairing) is associated with a
Berezinsky--Kosterlitz--Thouless (BKT) type transition~\cite{BKT}
is not ruled out.

Recently, it was found that the matter fields in the Abelian Higgs
model lead to a logarithmic attraction between monopoles and
antimonopoles~\cite{AnomalousMatter} which would explain the
formation of monopole--antimonopole bound states and string
breaking.  Adding massless
quarks also forces the Abelian monopoles and antimonopoles to form bound
states~\cite{Agasian:2001an}. Note that the origin of monopole
binding in the zero temperature case of the cAHM$_3$ is physically
different from the monopole binding observed at the finite
temperature phase transition in compact ($2+1$)--dimensional
pure QED~\cite{Binding,CIS12}.

We have recently studied the effect of finite temperature deconfinement
of ($2+1$)--dimensional cQED on the photon propagator in
Refs.~\cite{CIS3,CIS4}.We could demonstrate that the momentum behaviour of
the photon propagator in this theory is described, rather similar to
gluodynamics, by a Debye mass and by an anomalous dimension which both
vanish at the deconfinement transition. This mechanism could be clearly
attributed to pairing of magnetic monopoles. The monopole--antimonopole
plasma contribution is relatively easy to exhibit by explicit calculation
and can be eliminated by {\it monopole subtraction}from the ensemble of
gauge fields.

As for gluodynamics, which motivated our study of the cAHM$_3$,
numerical lattice results show that the propagator for all these gauges in
momentum space is less singular than $1/p^2$ in the immediate vicinity of
$p^2 = 0$. Recent investigations in the Landau gauge and in the Laplacian
gauge\footnote{In order to avoid the problem of Gribov copies~\cite{Gribov},
the alternative Laplacian gauge has also been used recently~\cite{deForcrand}
in Yang--Mills theory. However, principal questions of renormalizability
and transversality are yet unsolved in this case.} show that, beside the
suppression at $p^2 \to 0$, the propagator is enhanced at intermediate
momenta which can be characterized by an anomalous dimension~\cite{CurrentQCD}.
This enhancement of the Landau gauge propagator in $SU(2)$ gluodynamics has
been interpreted~\cite{Langfeld} by focusing on $P$-vortices appearing in
the maximal center gauge. Subtracting the vortices removes the enhancement
at intermediate momenta. The results for the propagator at zero momentum
are ranging from a finite~\cite{deForcrand} (Laplacian gauge) to a
strictly vanishing~\cite{Gribov,Gribov:numerical,Zwanziger} (Coulomb gauge)
value. The vanishing of the Landau gauge propagator at $p^2 = 0$, suggested
by considering the Faddeev--Popov mechanism, remains obscured so far in the
results, probably because the lattices are still too small.

In this paper we are going to investigate in which way the photon
propagator within cAHM$_3$ changes at the string breaking
crossover, turning from confinement to the symmetry--broken Higgs region.
As in our previous work we have chosen the propagator in the minimal
Landau gauge because this is the covariant gauge which has been adopted in
most of the investigations of the gauge boson propagators in
QCD~\cite{CurrentQCD,Langfeld} and QED~\cite{QEDpropagators}.  For the
behaviour at the string breaking crossover of  cAHM$_3$ we anticipate that a
confining propagator will change into a Yukawa--like propagator corresponding
to the onset of the Higgs mechanism.

The present paper is structured as follows. Next, in
Section~\ref{sec:model}, we will recall the model and the definition of
the photon propagator. The form of the fitting function and the
method of monopole subtraction are also introduced there. In
Section~\ref{sec:results} we will report the numerical results of
the present study and discuss the sensitivity with respect to the
Gribov copy problem. We conclude in Section~\ref{sec:conclusions}.

\section{The Model and the Definition of the Propagator}
\label{sec:model}

We consider the three--dimensional Abelian gauge model with compact gauge
fields $\theta_{x,\mu}$
living on links
and a fundamentally charged Higgs fields
$\Phi_x$ on sites. For simplicity we consider the London limit of
the model, which corresponds to an infinitely deep potential for
the Higgs field. Consequently, the radial part of the Higgs field,
$|\Phi_x|$, gets frozen.
With $\varphi_x = \arg \Phi_x$ we define the model by
the action
\beqn
S[\theta] = - \beta \sum_P \cos\theta_P
            - \kappa \sum_{x,\mu}
        \cos(\theta_{x,\mu} + \varphi_{x+\hat{\mu}} - \varphi_x) \,,
\label{eq:action}
\eeqn
where $\beta$ is the inverse gauge coupling squared,
$\kappa$ is the hopping parameter,
and $\theta_P$ is the plaquette
angle representing the curl of the link field $\theta_{x,\mu}$.

For the simulations we use a Monte Carlo algorithm similar to the
one described in Refs.~\cite{CIS3,CIS4,CIS5}.
The Higgs field angles have been
updated in alternating order with the gauge field angles.
In both cases, one $5$--hit Metropolis sweep together with $2$ microcanonical
sweeps constitute a total gauge or Higgs update.
Global updates of the gauge field have a negligible acceptance
rate for $\kappa \ne 0$. They have been discarded at all.
The numerical calculations have been carried out for fixed gauge
coupling, $\beta = 2.0$, on lattices of size $32^3$, as in
Ref.~\cite{CIS5}. For these parameters, the low--$\kappa$ region
(``confinement region'') and the high--$\kappa$ region (``Higgs
region'') of the phase diagram are separated by a crossover. The
crossover point --- observed as a position of the peak of the
susceptibility of the hopping term
$S_{GH} =   \sum_{x,\mu}
        \cos(\theta_{x,\mu} + \varphi_{x+\hat{\mu}} - \varphi_x)$
--- is located at $\kappa_c = 0.526(1)$.
We have considered from $O(200)$ to $O(400)$ independent configurations
(obtained after 10 subsequent updates) to measure the propagator.

The discussion of the photon propagator (and its various parts) is given in
lattice momentum space. Being always defined in the context of a specified
gauge, in our case the minimal Landau gauge
\beqn
\sum_{x,\mu} \cos(\theta^G_{x,\mu}) \rightarrow {\rm max}
\label{def:Landau_gauge}
\eeqn
with respect to gauge transformations $G$,
the propagator is written in terms of the Fourier transformed gauge potential,
\beqn
\tilde{A}_{{\vec k},\mu} =
\frac{1}{\sqrt{L^3}}
\sum\limits_{{\vec n}}
\exp \Bigl( 2 \pi i~\sum_{\nu=1}^{3} \frac{k_{\nu}
(~n_{\nu}+\frac{1}{2}\delta_{\nu\mu}~) } {L_{\nu}} \Bigr)
~A_{{\vec n}+\frac{1}{2}{\vec \mu},\mu} \,,
\label{def:fourier_transformation}
\eeqn
which is a sum over a set of points ${\vec x}={\vec n}+\frac{1}{2}{\vec \mu}$,
the midpoints of the links in $\mu$ direction, which form the support of
$A_{{\vec x},\mu}$ on the lattice.
${\vec n}$ denotes the lattice sites (nodes) with integer Cartesian coordinates.
The propagator is the gauge--fixed ensemble average
of the following bilinear in $\tilde{A}$,
\beqn
D_{\mu\nu}({\vec p}) = \langle \tilde{A}_{ {\vec k},\mu}
                               \tilde{A}_{-{\vec k},\nu} \rangle \,,
\label{def:propagator}
\eeqn
where the lattice momenta ${\vec p}$ on the left hand side of
(\ref{def:propagator}) are related to the integer valued
Fourier momenta ${\vec k}$ by the expression
($a$ is the lattice spacing):
\beqn
p_\mu(k_\mu)=  \frac{2}{a} \sin \frac{ \pi k_\mu}{L_\mu}
\,, \quad  k_\mu=0, \pm 1,..., \pm \frac{L_\mu}{2} \,.
\label{def:momenta}
\eeqn
The lattice equivalent of $p^2 = {\vec p}^{\,2}$ is in $3$ dimensions
\beqn
p^2({\vec k})=
\frac{4}{a^2} \sum_{\mu=1}^3 \left(\sin \frac{\pi k_\mu}{L_\mu}
\right)^2  \, .
\label{def:momentum_squared}
\eeqn

{}For this paper, we decided to identify the gauge field
$A_{{\vec x},\mu}$ in terms of the sine function of the link angle
(with $g_3^2= 1/(a \beta)$)
\beqn
A_{\vec n+\frac{1}{2}{\vec \mu},\mu}
  =\frac{1}{g_3 a} \sin \theta_{\vec n,\mu} \,.
\label{def:sine}
\eeqn
The corresponding propagator, called $D^{\sin}_{\mu\nu}$ in Ref.~\cite{CIS4},
will be called simply $D_{\mu\nu}$ for brevity.
The extraction of the Fourier transformed gauge field (and of its regular
and singular components) has to be performed after the original gauge field
configuration has been put into the minimal Landau gauge.
The procedures employed for gauge fixing in the present context have
been described at length in Ref.~\cite{CIS4}.

At zero temperature, for perfect Euclidean rotational invariance,
the continuum propagator would be expressible by functions of $p^2$.
The most general tensor structure is then the following one including
two scalar functions of $p^2$,
\beqn
D_{\mu\nu}({\vec p})=
P_{\mu\nu}({\vec p})~D(p^2) + \frac{p_\mu p_\nu}{p^2} \frac{F(p^2)}{p^2}
\label{def:tensor_structure}
\eeqn
with the three--dimensional  transverse projection operator
\beqn
P_{\mu\nu}({\vec p}) = \delta_{\mu\nu}- \frac{p_\mu p_\nu}{p^2} \, .
\label{def:transverse_proj}
\eeqn
The two structure functions $D(p^2)$ and $F(p^2)$
can be extracted by projection, on the lattice from $D_{\mu\nu}({\vec p})$
according to (\ref{def:propagator}), as
\beqn
F(p^2)=  \sum\limits_{\mu,\nu=1}^{3} p_\mu~ D_{\mu\nu}({\vec p})~ p_\nu
\label{def:F_project_1}
\eeqn
and
\beqn
p^2~D(p^2)= \frac{1}{2} \sum\limits_{\mu,\nu=1}^{3}
P_{\mu\nu}({\vec p})~D_{\mu\nu}({\vec p}) \, .
\eeqn
They are approximately rotationally
invariant, {\it i.e.} individual momenta ${\vec p}$ might slightly differ
in the function values $D$ or $F$ they provide, even if they have the
same $p^2$.
Dense, in $p^2$ nearby data points may scatter rather than be forming
a smooth function of $p^2$.

In practice, using these definitions, we extract at first the function
$F(p^2)$ from Eq.~(\ref{def:F_project_1}). In the sum the
imaginary parts of non-diagonal $D_{\mu\nu}$ cancel.
Then, $D(p^2)$ is obtained through
\beqn
 D(p^2)= \frac{1}{2}\left\{\Big[ D_{11}({\vec p})
                                 + D_{22}({\vec p})
                                 + D_{33}({\vec p}) \Big]
            - \frac{F(p^2)}{p^2} \right\} \, .
\label{def:D_project_2}
\eeqn

For exactly fulfilled Landau gauge $F(p^2) \equiv 0$.
On the lattice, in the case of the sine--definition used
for $A_{{\vec x},\mu}$ [Eq.~(\ref{def:sine})],
this is actually the case as soon as one of the local maxima of
(\ref{def:Landau_gauge}) (Gribov copies) is reached, with an
accuracy which directly reflects the precision at stopping of the
gauge fixing iterations.

The effect of monopoles (singular fields) can be distinguished from
that of the regular (``photon'') fields using
the splitting of the gauge field  angles $\theta_{x, \mu}$ into a regular
and a monopole part following
Refs.~\cite{PhMon,CIS3,CIS4}. In the notation of lattice differential
forms this can be written as:
\beqn
\theta = \theta^{\reg} + \theta^{\mon}\,, \quad
  \theta^{\mon} = 2 \pi \Delta^{-1} \delta p[j]\,,
  \label{eq:theta-decomp}
\eeqn
where $\Delta^{-1}$ is the inverse lattice Laplacian and the
0-form $\dual j \in \Z$ is nonvanishing on the sites of the dual
lattice occupied by monopoles and antimonopoles. The 1-form  $\dual p[j]$
corresponds to Dirac strings (living on the links of the dual
lattice) which connect monopoles with antimonopoles,
$\delta \dual p[j] = \dual j$. For any Monte Carlo configuration,
we have fixed the gauge, then located the Dirac strings, $p[j]\ne0$,
and constructed the monopole part $\theta^{\mon}$ of the gauge field
according to the last equation in \eq{eq:theta-decomp}. The regular
photon field\footnote{In principle, the regular part of the links could
have been reconstructed as well, without recurrence to the singular part.
In this case the regular propagator would become {\it completely}
independent of the number of Gribov copies $N_G$ (see the discussion
below).}
is taken just as the complement to the monopole part according to the
first equation of \eq{eq:theta-decomp}.

The regular and the singular parts of the gauge field
contribute to the propagator as follows: the total propagator
decomposes like $D = D^{\reg} + D^{\mon} + D^{\mix}$, where
$D^{\mix}$ represents the mixed contribution from regular
{\it and} singular fields.
An analogous decomposition is valid for the longitudinal structure
function $F$ which vanishes to a good accuracy for the {\it total} propagator.
In this paper we are interested mainly
in the regular part and in the total photon propagator.
To describe the propagators quantitatively we have fitted the
total propagator using the following function:
\beqn
  D(p^2) = \frac{Z~m^{2 \alpha}}
  { \beta~(p^{2 (1+\alpha)} + m^{2 (1+ \alpha)}) } + C\,,
  \label{eq:fit:total}
\eeqn
where $Z$, $\alpha$, $m$ and $C$ are the fitting parameters.
This fit has been successfully used to describe the propagators
of the finite and zero--temperature compact U(1) gauge
model~\cite{CIS3,CIS4} in $2+1$ or $3$ dimensions, respectively.
The form is similar to some of Refs.~\cite{CurrentQCD} where the
propagator in gluodynamics has been studied. The meaning of the
fitting parameters in Eq.~\eq{eq:fit:total} is as follows: $Z$ is
the renormalization of the photon wavefunction, $\alpha$ is the
anomalous dimension, $m$ is a mass parameter. As shown in
Ref.~\cite{CIS4}, in cQED$_3$ this mass parameter coincides with
the Polyakov prediction~\cite{Polyakov} for the Debye mass,
generated by the monopole--antimonopole plasma.
The parameter $C$ corresponds to
a $\delta$--like interaction in the coordinate space and,
consequently, is irrelevant for long--range physics.

The regular (or ``photon'') part of the propagator, $D^{\reg}$, has
been fitted by the Yukawa form (with an additional contact term):
\beqn
  D^{\reg}(p^2) = \frac{Z_{\reg}}{ \beta~(p^2 + m_{\reg}^2) } + C_{\reg} \,.
  \label{eq:fit:photon}
\eeqn
The fits together with the appropriate propagator data will
be presented in the next section.

\section{Numerical Results}
\label{sec:results}

\begin{figure*}[!htb]
\begin{center}
\begin{tabular}{cc}
\epsfxsize=7.5cm \epsffile{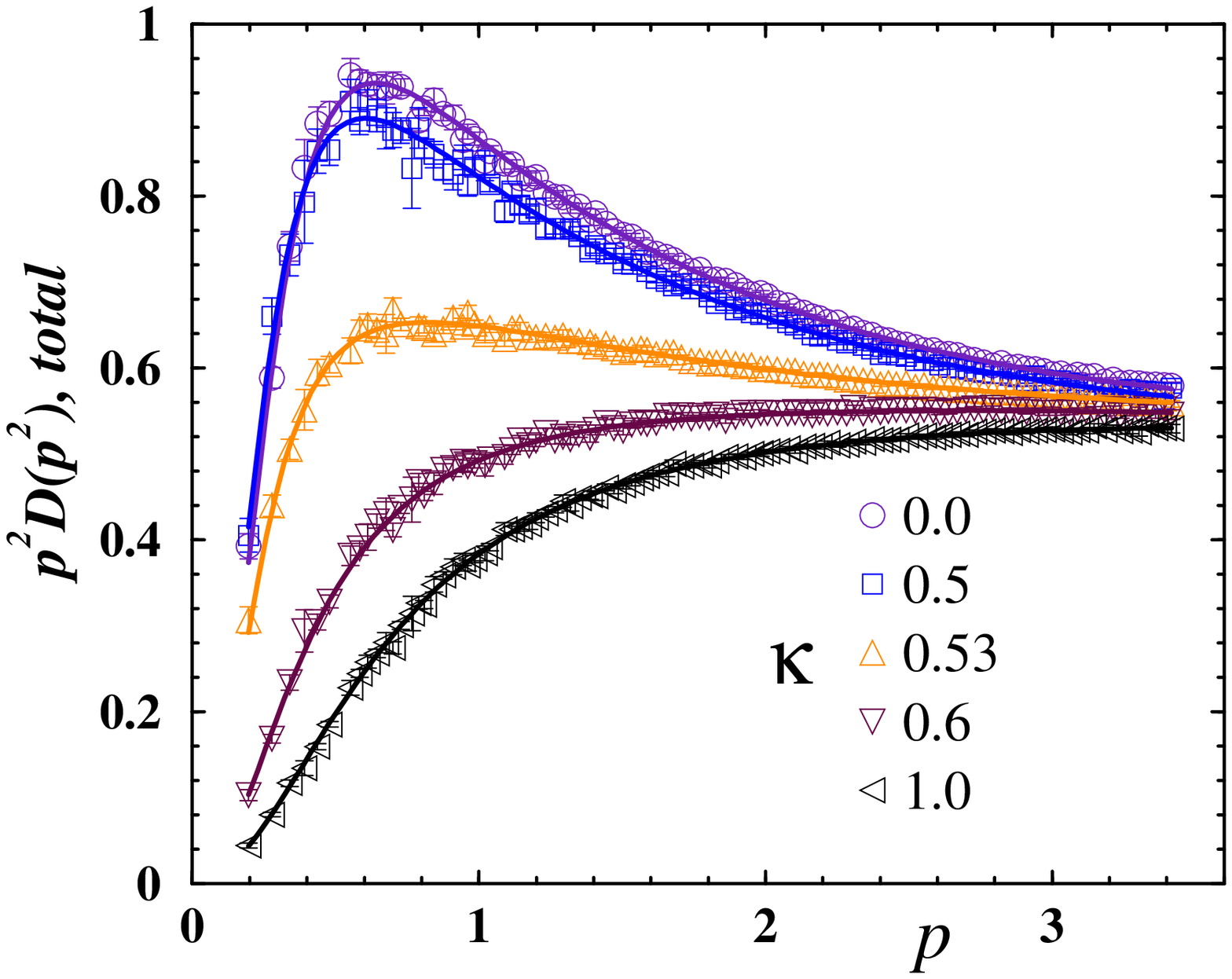}  &
\epsfxsize=7.5cm \epsffile{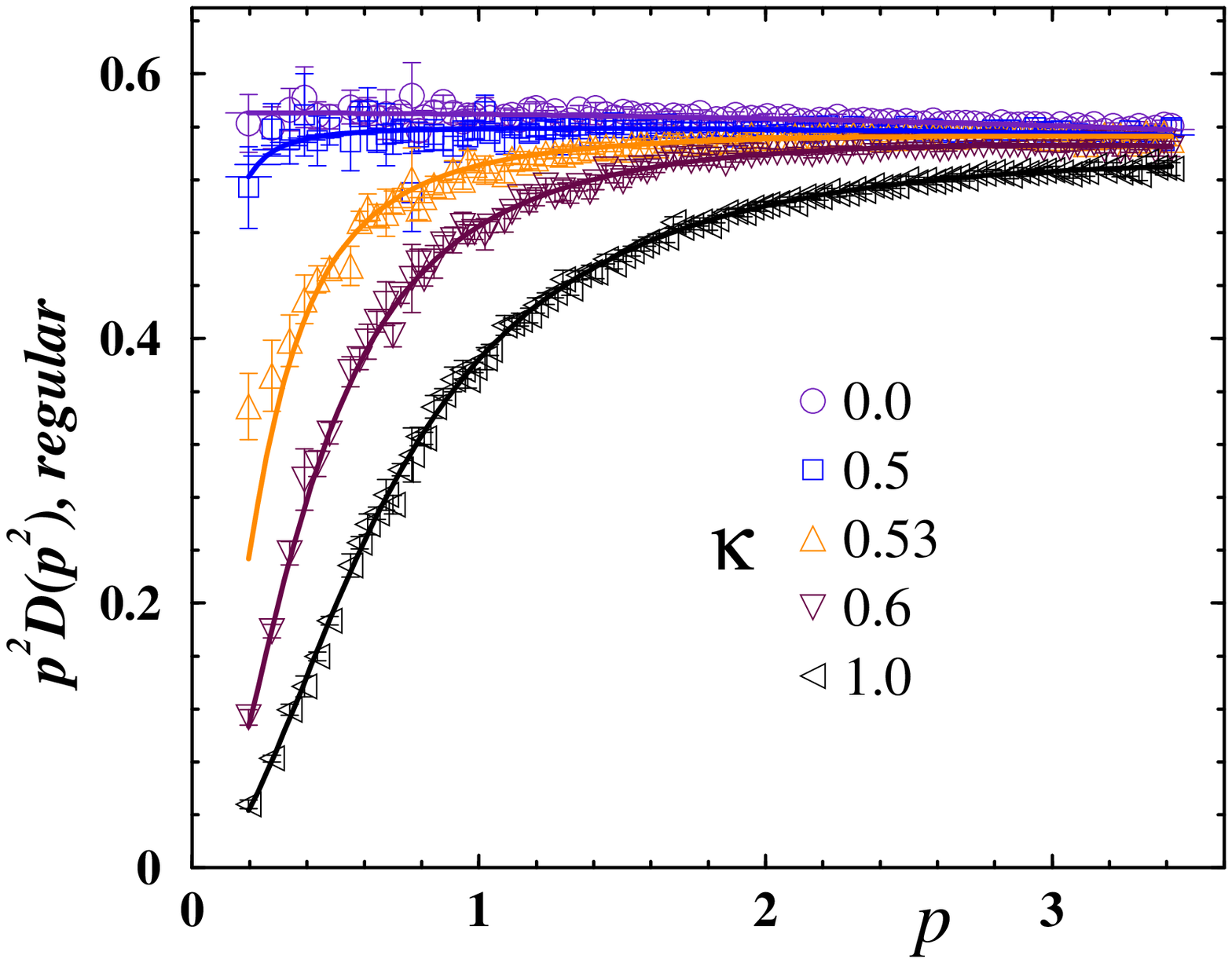} \\
  (a) & \hspace{5mm} (b)
\end{tabular}
\end{center}
\caption{
Momentum dependence of the total propagator (a) and its regular
part (b) and fitted curves}
\label{fig:fitting}
\end{figure*}

We begin with the typical shape of the total photon propagator and
its regular part, respectively, at fixed $\beta = 2.0$.
The total photon propagator $D$, multiplied by $p^2$, is shown in
Figure~\ref{fig:fitting}(a)
for various values of $\kappa$ as a function of $|{\vec p}|$.
We also present the fits of the data by the function~\eq{eq:fit:total}.
One clearly observes that the total propagator is well described by the
fitting function in each particular case. The ``infrared'' enhancement
at intermediate momenta disappears with increasing $\kappa$. In
contrast, the suppression at very small $|{\vec p}|$ remains at
all $\kappa$ values. We can conclude that the total photon
propagator is less singular in the infrared than a free one.
According to our fitting function~\eq{eq:fit:total} the propagator
is finite at $p^2=0$.

For comparison, the regular photon propagator part
$D^{\reg}$, multiplied by $p^2$, is shown in Figure~\ref{fig:fitting}(b)
for various values of $\kappa$. One can see that the behaviour
of $D^{\reg} \propto 1/p^2$ is only observed in the confinement region; in
the Higgs region the propagator becomes massive. The fits
using function~\eq{eq:fit:photon} work very well everywhere
except for smallest non--zero momenta in the closest vicinity
of the crossover point.

Note that the data for both the
total propagator and the regular part have been
averaged over lattice momenta corresponding to the same $p^2$
before fitting (as in Refs.~\cite{CIS3,CIS4}).

\begin{figure*}[!htb]
\begin{center}
\begin{tabular}{rr}
\epsfxsize=6.8cm \epsffile{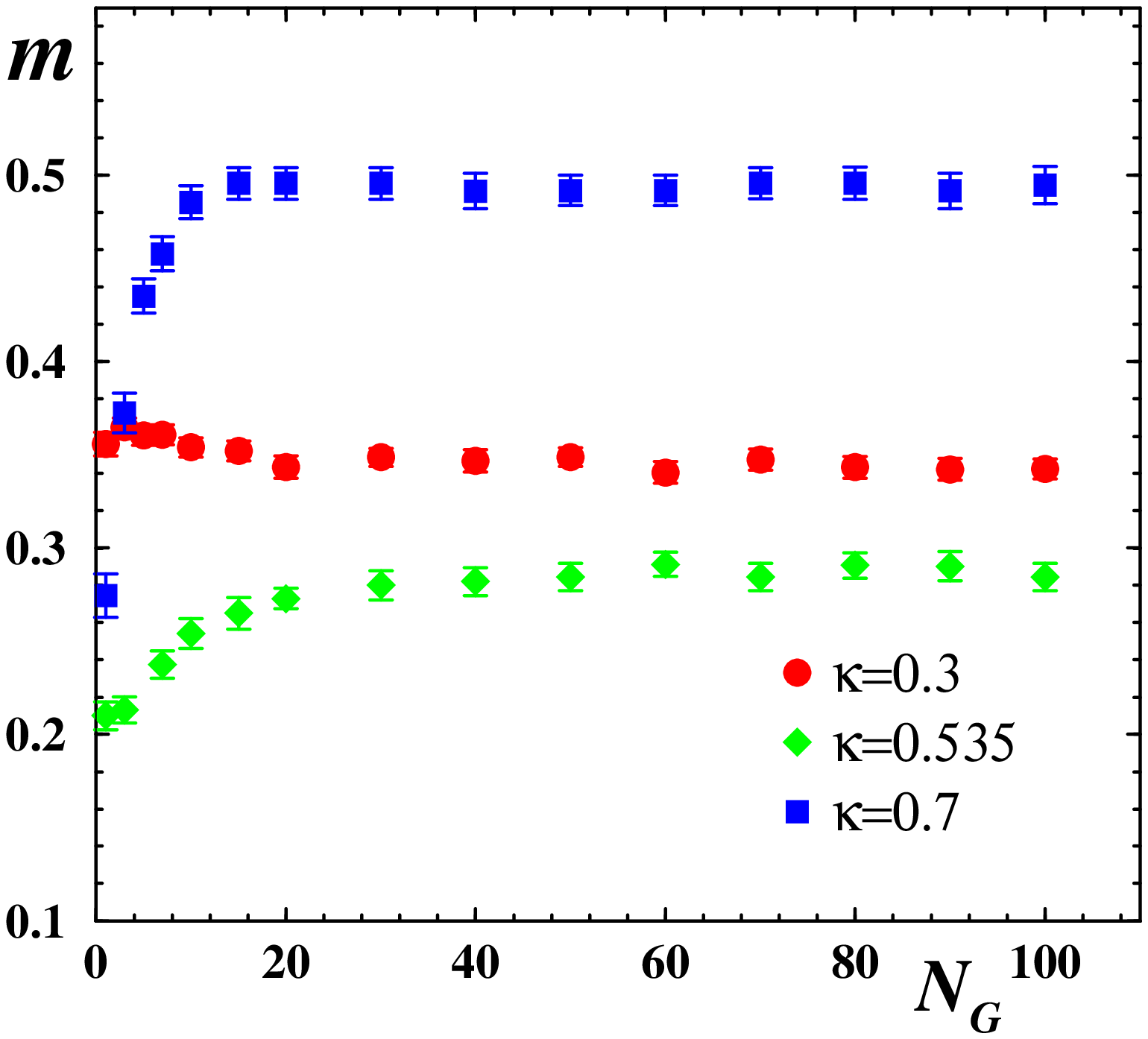}  &
\epsfxsize=7.0cm \epsffile{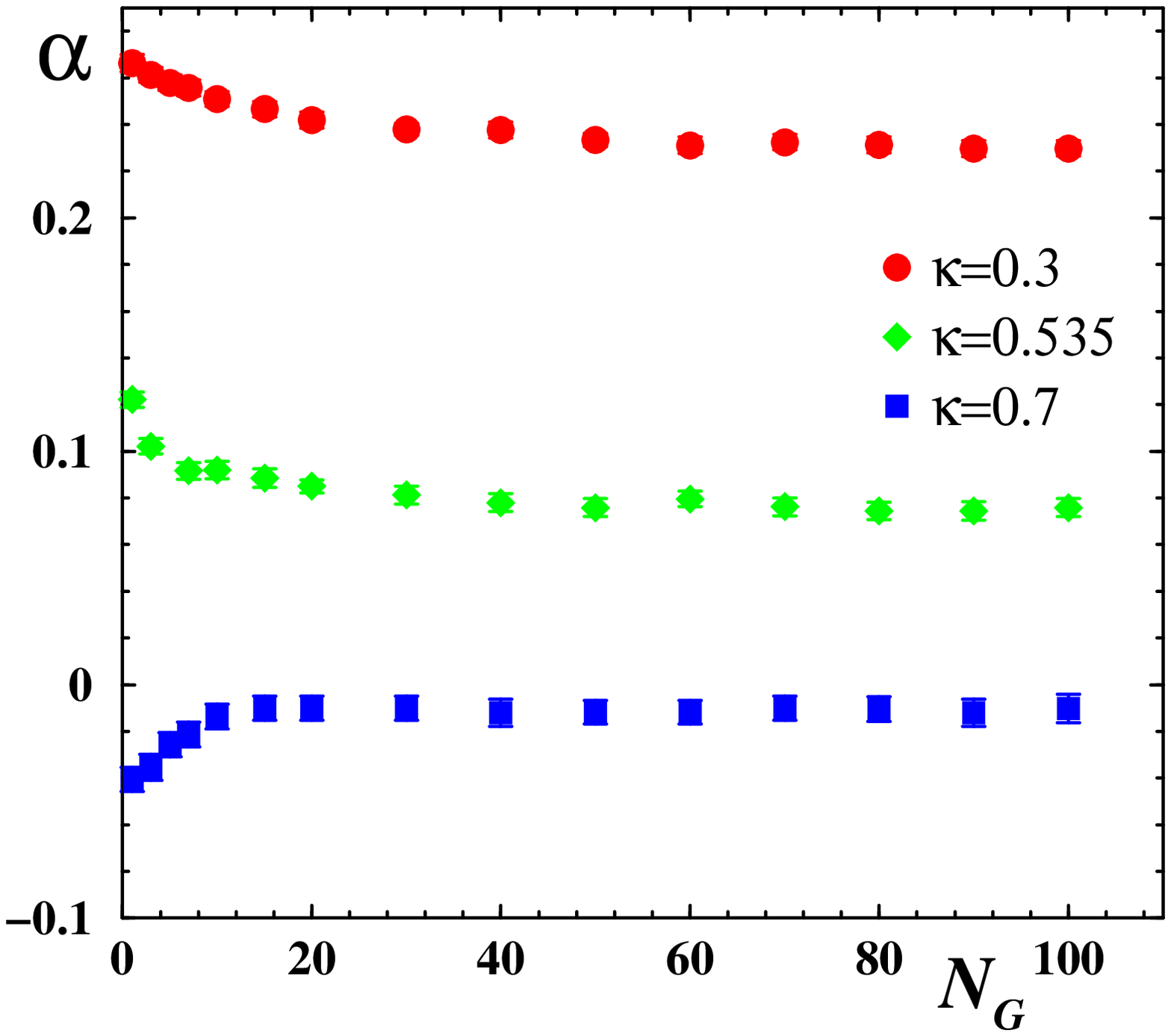} \\
\epsfxsize=7.0cm \epsfxsize=6.9cm \epsffile{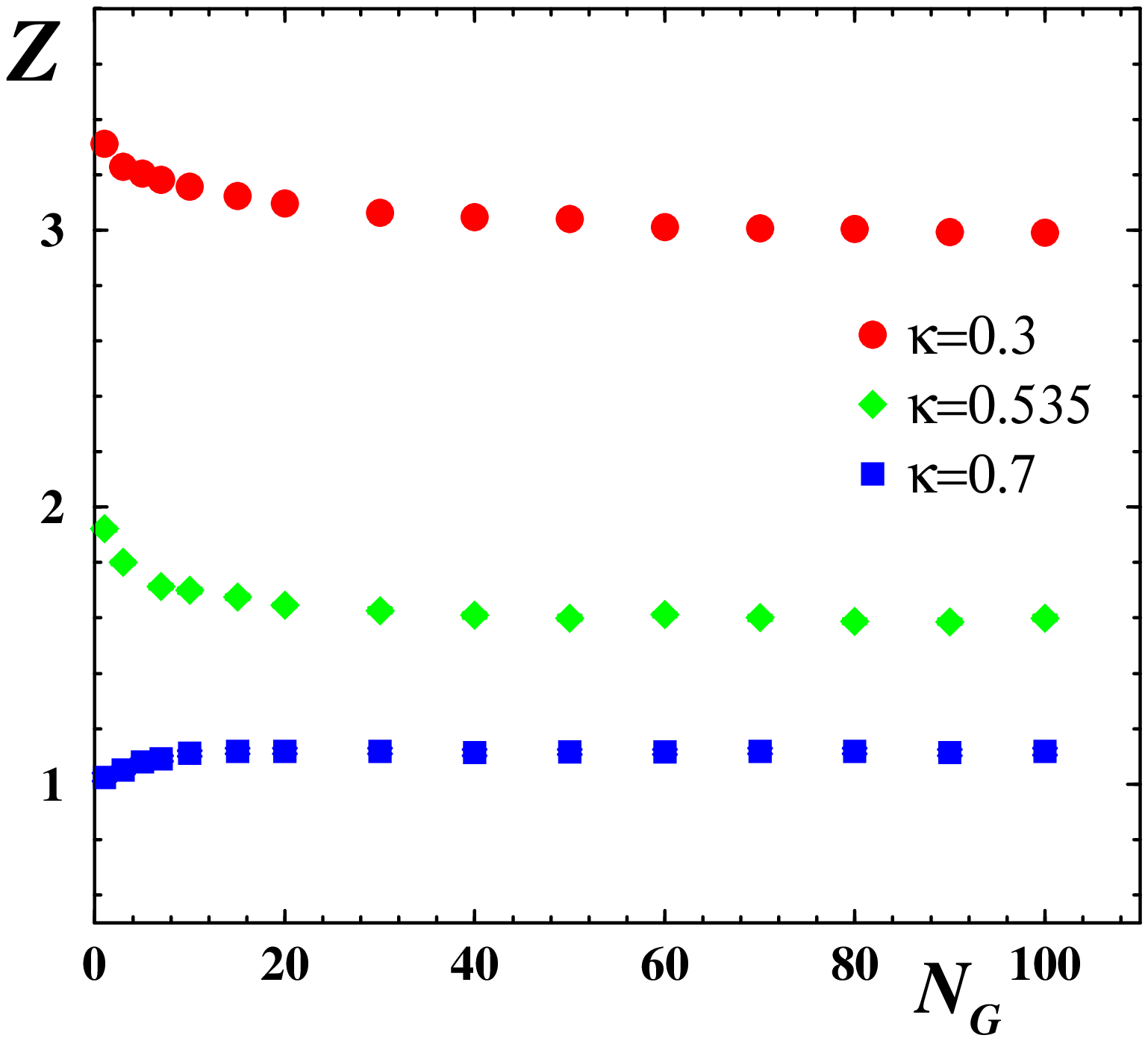}  &
\epsfxsize=7.4cm \epsffile{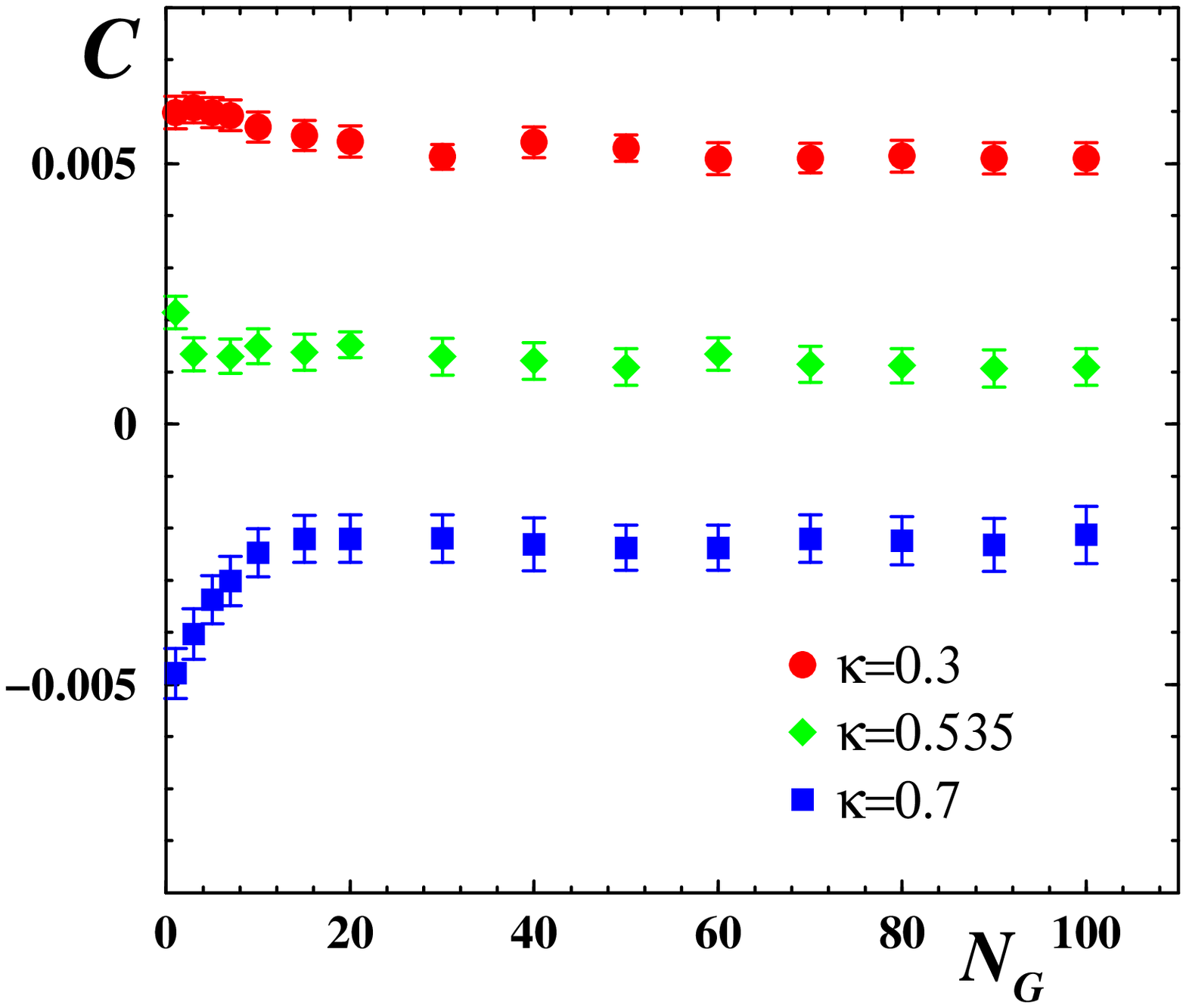}
\end{tabular}
\end{center}
\caption{Dependence of the fit parameters $m$, $\alpha$, $Z$ and $C$
on the number $N_G$ of Gribov copies.}
\label{fig:Gribov}
\end{figure*}

{}From the fits we have obtained the characterizing fit parameters
for the transverse photon propagator $D$ in gauge fixed ensembles
with various numbers ($N_G \leq 100$) of Gribov copies. This means
that, after applying the gauge fixing algorithm to $N_G$ random
gauge copies of the original Monte Carlo gauge field, only the
configuration with the maximal value of the gauge fixing
functional (\ref{def:Landau_gauge}) is kept for evaluation.
In Fig.~\ref{fig:Gribov}
we show the dependence of the fit parameters
for the total photon propagator on the number of Gribov gauge copies.
Three typical $\kappa$ values are used: $\kappa=0.3$ in the
confinement region, $\kappa=0.535$ very near the crossover and
$\kappa=0.7$ in the Higgs region. While a moderate number $N_G
\approx 30$ seems to be sufficient for convergence of all
parameters at all $\kappa$ values, there is no clear tendency
between different $\kappa$'s.
The parameters describing the regular propagator $D^{\reg}$ converge
within the first few ($N_G \approx 5$) copies.

\begin{figure*}[!htb]
  \begin{center}
  \begin{tabular}{cc}
  \epsfxsize=7.0cm \epsffile{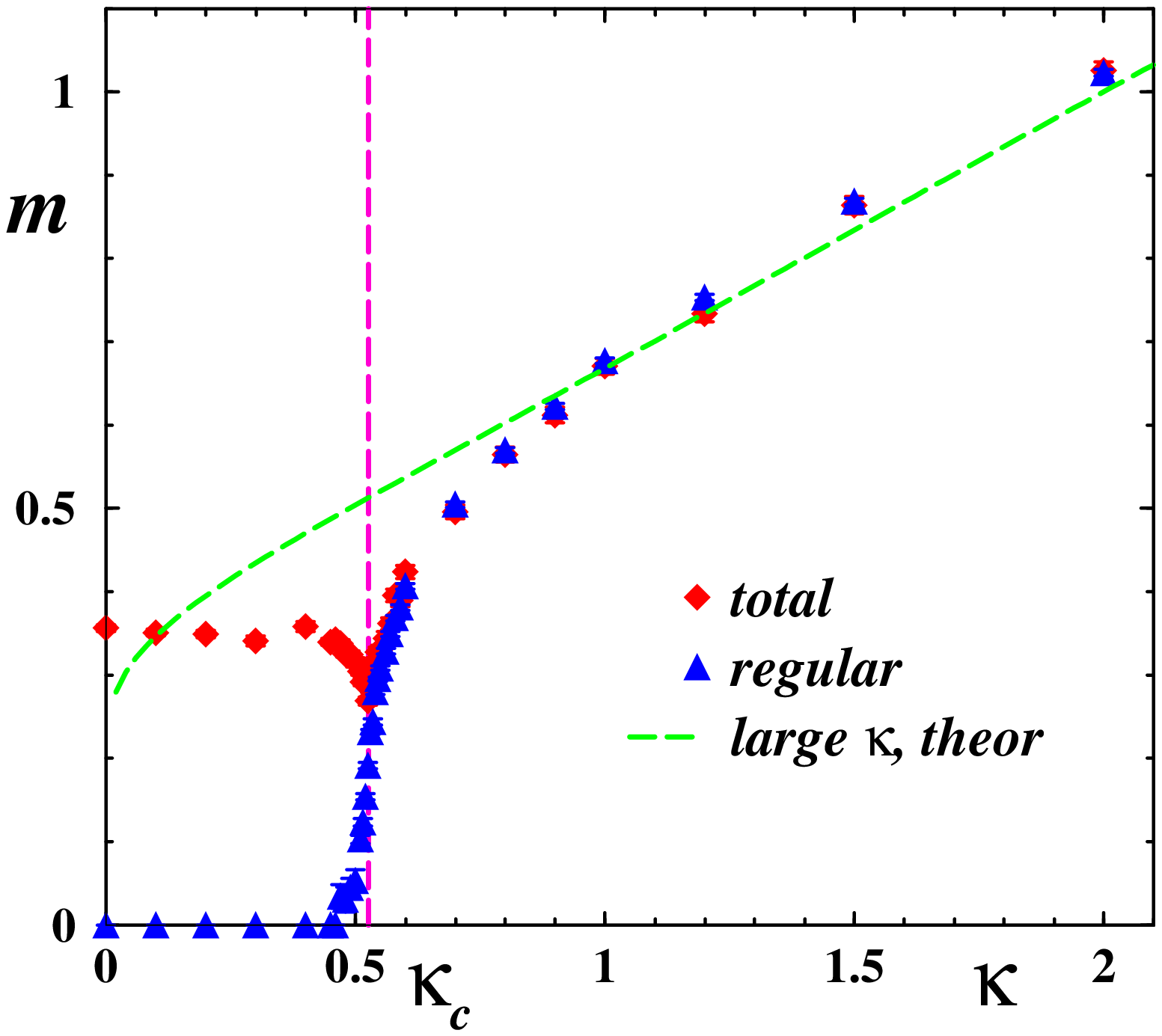}  & \hspace{5mm}
  \epsfxsize=7.0cm \epsffile{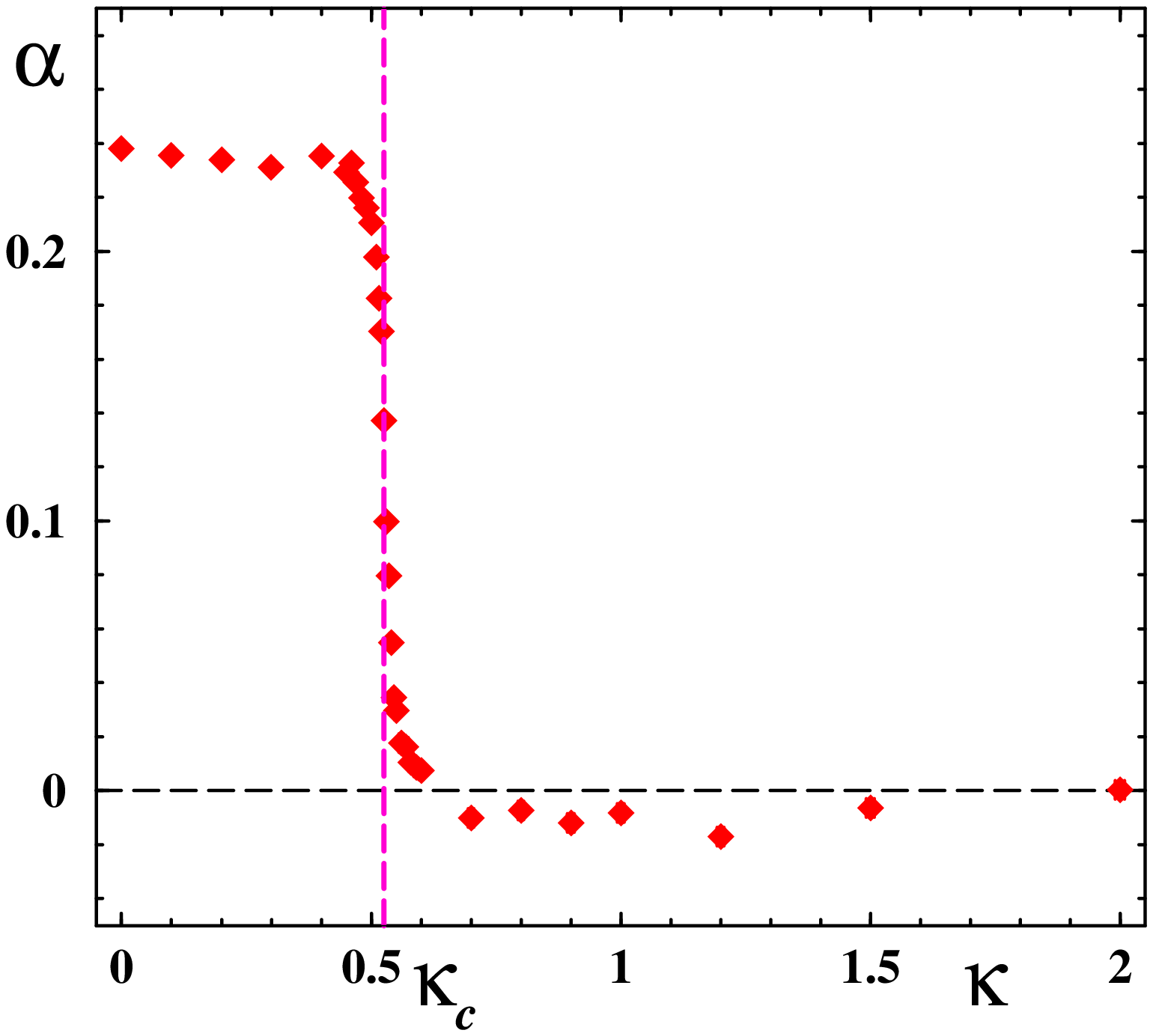} \\
  (a) & \hspace{5mm} (b)\\
  \epsfxsize=7.0cm \epsffile{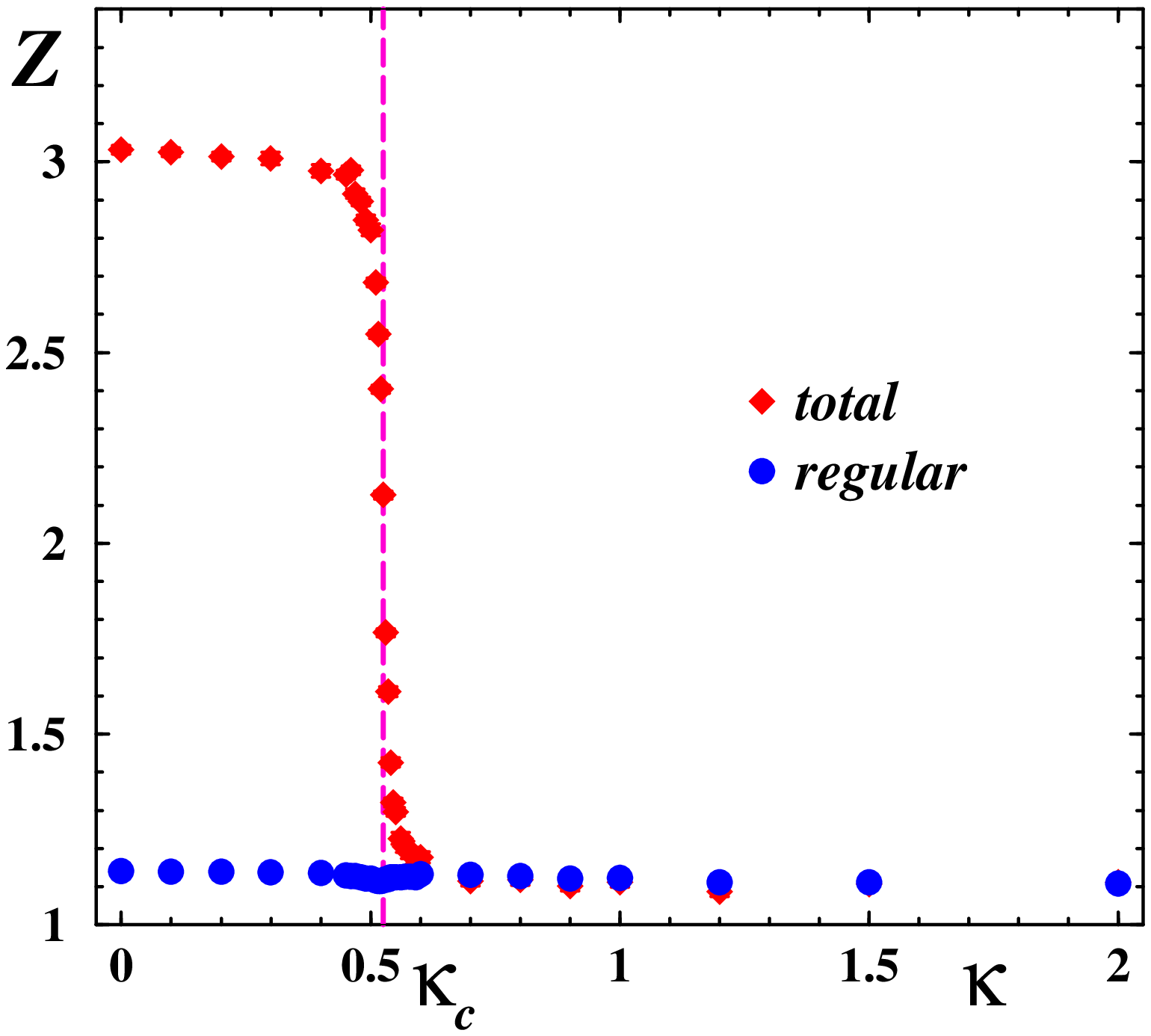}  & \hspace{5mm}
  \hspace {-8mm}
  \epsfxsize=7.7cm \epsffile{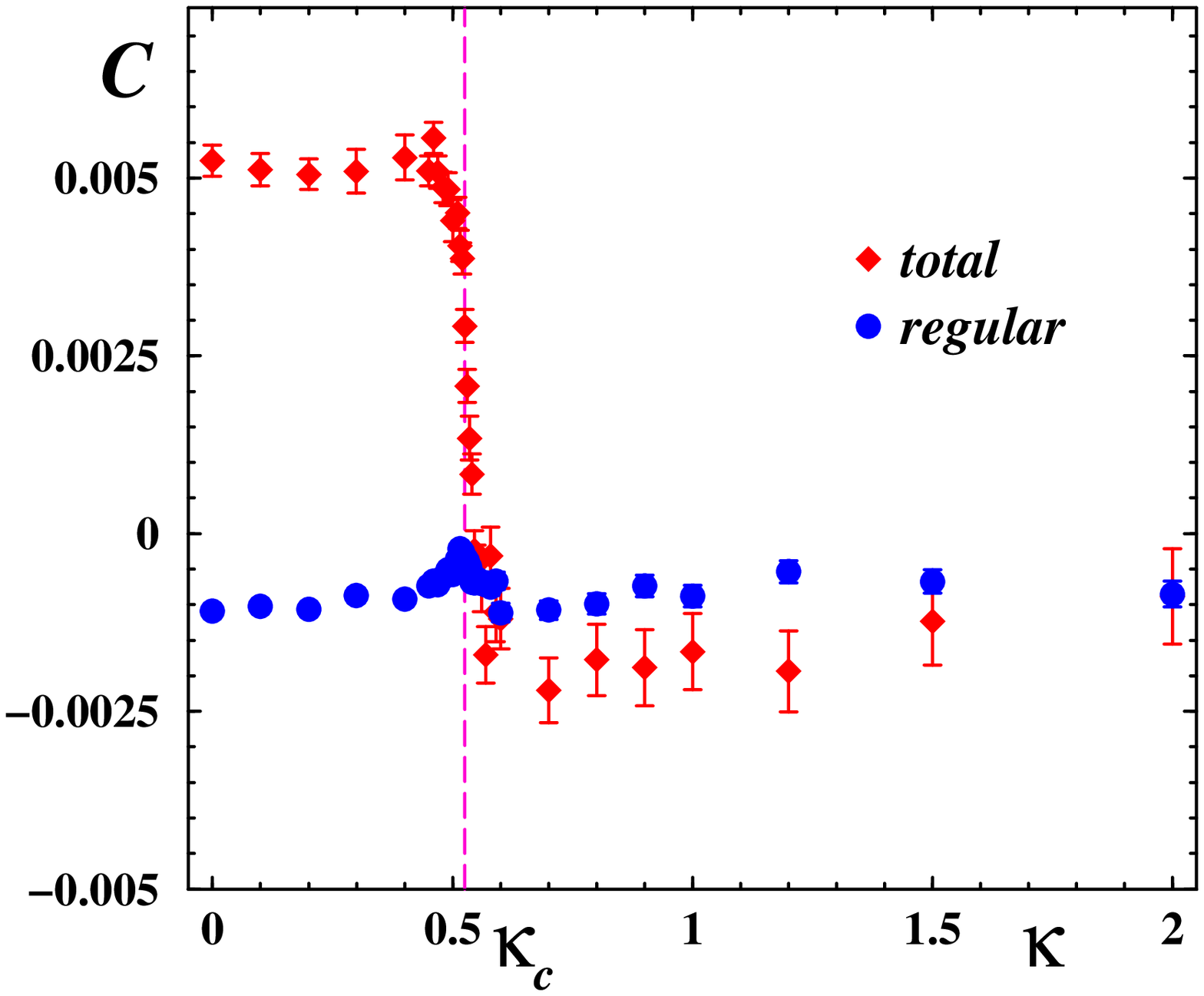} \\
  (c) & \hspace{5mm} (d) \\
  \end{tabular}
  \end{center}
  \vspace{-5mm}
  \caption{Fit parameters for the propagators at
  $\beta=2.0$ as function of $\kappa$:
  (a) $m$ and $m_{\reg}$ together with the analytical prediction for large $\kappa$;
  (b) $\alpha$;
  (c) $Z$ and $Z_{\reg}$ and
  (d) $C$.
  The crossover point is denoted by the vertical dotted line.}
  \label{fig:bestfits}
\end{figure*}

To keep the influence of $N_G$ on the propagator negligible
we have used in the final measurements $N_G=60$. The
resulting fit parameters of interest are presented in
Fig.~\ref{fig:bestfits} as functions of $\kappa$.
Fig.~\ref{fig:bestfits}(a) depicts the mass taken from the
fit~\eq{eq:fit:photon} for the regular part of the propagator
("regular mass" $m_{\reg}$), and from the fit~\eq{eq:fit:total}
for the total propagator ("total mass", $m$). The dashed line represents
the expected behaviour of the mass in the limit of large $\kappa$
and $\beta$ couplings:
\beqn
  m_{\mathrm{th}}(\beta,\kappa) =
  \sqrt{\frac{\kappa_V(\kappa)}{\beta_V(\beta)}}\,,
  \label{eq:m:theor}
\eeqn
where $\beta_V$ and $\kappa_V$ are the Villain couplings corresponding
to the Wilson couplings $\beta$ and $\kappa$.
This parameterization is analogous to Ref.~\cite{BMK}, with
\beqn
  \beta_V(\beta) = {\Bigl[ 2 \log
  \frac{I_0(\beta)}{I_1(\beta)}\Bigr]}^{-1}\,,\quad
  \kappa_V(\kappa) = {\Bigl[ 2 \log
  \frac{I_0(\kappa)}{I_1(\kappa)}\Bigr]}^{-1}\,.
  \label{eq:BetaV}
\eeqn
Note, that although $\beta=2.0$ (used in our calculations) is not
very large, the first relation in Eq.~\eq{eq:BetaV} works with an
accuracy of a few percent already at $\beta \sim 1$ according to
the calculation~\cite{BMK} of the critical coupling in $4D$ cQED.

We observe that both the total and regular masses coincide with
each other on the Higgs side of the crossover, and there they are
very close to the prediction (\ref{eq:m:theor},\ref{eq:BetaV}),
the immediate vicinity of the crossover point excluded. This
observation can be easily understood.
Two different sources contribute to the gauge boson mass:
one is arising nonperturbatively from the monopoles
(the Debye screening mass) while the other is due to the explicit
presence of the mass term in the action~\eq{eq:action} of the
model. The Debye mass generation works in
the case of the monopole--antimonopole plasma~\cite{Polyakov} and
it is obviously absent in the magnetic dipole gas~\cite{DipoleGas}.
Since at large $\kappa$ monopoles and antimonopoles are bound into
dipoles~\cite{AnomalousMatter,CIS5},
the Debye contribution to the mass disappears, and the gauge boson
mass is exclusively given by Eq.~\eq{eq:m:theor}.
As can be seen from Fig.~\ref{fig:bestfits}(a), at small $\kappa$
values the total mass $m$ is close to the Debye mass of pure cQED$_3$
({\it i.e.} to the mass value at $\kappa=0$) because the effects of the
mass term in the action~\eq{eq:action} are small. Moreover, in
accordance with our expectations, the regular mass $m_{\reg}$ vanishes.

In the region very close to the crossover, $\kappa \approx \kappa_c$,
the mass $m$ shows a minimum caused, as one
could guess, by the interference between perturbative mass and Debye
mass effects. Indeed, as $\kappa$ tends to $\kappa_c$,
the Debye mass gets smaller since
the density of the monopole--antimonopole plasma drops rapidly.
One the other hand, the perturbative mass term becomes more significant.
The interplay of these two tendencies results in the noticed minimum at
$\kappa \approx \kappa_c$.

In Fig.~\ref{fig:bestfits}(b) we show the anomalous dimension
$\alpha$ for the total propagator\footnote{Recall, that the anomalous
dimension for the regular part of the propagator is zero,
Eq.~\eq{eq:fit:photon}.} as a function of the hopping parameter
$\kappa$. The anomalous dimension is
non--vanishing in the confined region and it turns to trivial values
at the crossover\footnote{The 4-parameter fits lead even to small
negative $\alpha$ in the region $\kappa \sim 1$. However, fitting
there the propagators with {\it fixed} $\alpha=0$, the obtained masses
coincide within errors with those of the unconstrained fits.
So we associate this behaviour with statistical fluctuations.},
$\alpha \rightarrow 0$. This behaviour can be compared with
our studies of compact QED$_{2+1}$ at finite temperature~\cite{CIS3,CIS4}:
in the monopole--antimonopole plasma phase (corresponding to the
confinement side) $\alpha >0$ while in the magnetic dipole phase
(corresponding to the Higgs side) $\alpha=0$. Thus, in cAHM the effect
of the monopole pairing on the propagator is the same as in cQED:
the anomalous dimension gets close to zero in the Higgs phase dominated
by the magnetic dipole gas.

Similarly to the anomalous dimension, the effect of the monopole
pairing on the renormalization parameter $Z$ of the photon
wavefunction in cAHM, shown in Figure~\ref{fig:bestfits}(c), is
remarkably similar to the cQED case
observed\footnote{ Due to an error, the $Z$ factors in Figs. 5c
and 7c of Ref.~\cite{CIS4} need to be corrected by a factor
$m^\alpha$ before they can be compared with the $Z$ given here at
$\kappa=0$.}
in Refs.~\cite{CIS3,CIS4}. The total photon factor
$Z$ suddenly drops at the crossover point, $\kappa = \kappa_c$
while the regular factor, $Z_{\reg}$, is almost insensitive with
respect to the crossover ($Z \rightarrow Z_{\reg} \approx 1$ for
all $\kappa$ values).
The small contact term parameter  of the total propagator, $C$,
changes its sign at the crossover, Figure~\ref{fig:bestfits}(d).
The small and negative contribution of $C_\reg$ in the confinement
is responsible for the tiny decrease of the massless
$p^2 D^\reg$ with $p^2$ [seen at zero and small $\kappa$ in
Fig.~\ref{fig:fitting}(b)] .
\begin{figure*}[!htb]
  \begin{center}
  \epsfxsize=7.3cm \epsffile{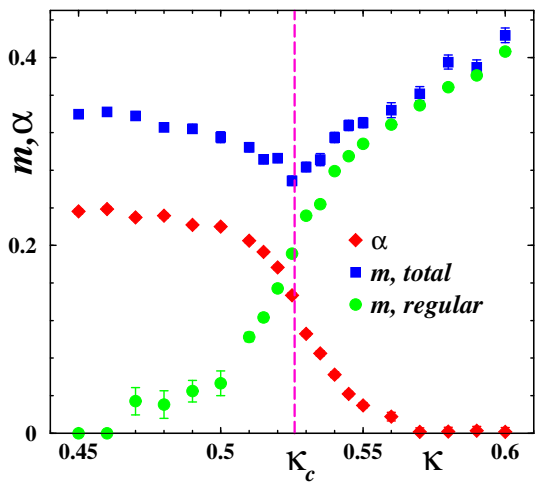}
  \end{center}
\caption{Zoom to masses and anomalous dimension in the crossover region.}
  \label{fig:zoom}
\end{figure*}
A zoom to the crossover region is shown in
Fig.~\ref{fig:zoom}, where all effects described above for
$m$, $m_{\reg}$ and $\alpha$ are seen more clearly together.

\section{Conclusions}
\label{sec:conclusions}

In our numerical studies we found that the gauge boson propagator
in the London limit of the three--dimensional compact Abelian
Higgs model possesses a non--zero anomalous dimension below the
string breaking crossover. The effects of the matter fields on the
propagator are remarkably similar to the finite--temperature
deconfining effects in the pure gauge compact $U(1)$ model. Both
in the Higgs (deconfinement) region of cAHM and in the
deconfinement (high temperature) phase of cQED the anomalous
dimension of the propagator vanishes. In the confining regions of
both models the anomalous dimension is non--vanishing. The
positive anomalous dimension is due to presence of the
monopole--antimonopole gas in the plasma state~\cite{CIS3}.
As we move towards the deconfinement phase $\alpha$ decreases and
becomes zero when the monopole--antimonopole plasma turns into a
magnetic dipole gas. However, the origin of the pairing phenomenon
in both models is different: in cAHM the monopole pairing is
caused by the matter fields~\cite{AnomalousMatter} while in cQED
the monopoles form the bound states due to temperature effects in
the monopole--antimonopole interaction.

For the limit of cAHM with frozen radial Higgs degrees
of freedom we have found that the anomalous dimension for the
gauge field does not become
clearly negative together with the onset of string breaking.
However, if radial fluctuations of the Higgs field would be allowed,
the emergence of a negative anomalous dimension cannot be ruled
out~\cite{Herbut,AnomalousMatter}.

The mass in the gauge boson propagator closely follows the tree
level expectations on the Higgs side of the transition and simply
corresponds to a massive Yukawa propagator practically unaffected
by remaining dipoles in the vacuum. In the confinement region the
mass related to the total gauge field is of Debye type while the
mass of the regular or photon part of the gauge field degrees of
freedom remains massless.

\section*{Acknowledgements}
M.~N.~Ch. is supported by the JSPS Grant No. P01023.
E.--M.~I. acknowledges useful discussions with L. von Smekal.

\end{document}